\documentclass[8.5pt,twoside,twocolumn]{article}
\usepackage[super,sort&compress,comma]{natbib}
\usepackage{mhchem}
\usepackage{sectsty}
\usepackage{balance}
\usepackage{color}

\usepackage{graphicx} 
\usepackage{lastpage}
\usepackage[format=plain,justification=raggedright,singlelinecheck=false,font=small,labelfont=bf,labelsep=space]{caption}
\usepackage{fancyhdr}
\begin{document}

\thispagestyle{plain}
\fancypagestyle{plain}{
\fancyhead[L]{\includegraphics[height=8pt]{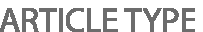}}
\fancyhead[C]{\hspace{-1cm}\includegraphics[height=20pt]{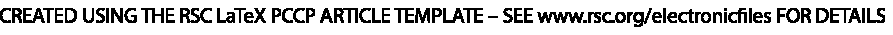}}
\fancyhead[R]{\includegraphics[height=10pt]{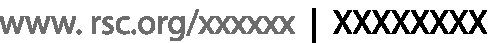}\vspace{-0.2cm}}
\renewcommand{\headrulewidth}{1pt}}
\renewcommand{\thefootnote}{\fnsymbol{footnote}}
\renewcommand\footnoterule{\vspace*{1pt}%
\hrule width 3.4in height 0.4pt \vspace*{5pt}}
\setcounter{secnumdepth}{5}

\makeatletter
\def\subsubsection{\@startsection{subsubsection}{3}{10pt}{-1.25ex plus -1ex minus -.1ex}{0ex plus 0ex}{\normalsize\bf}}
\def\paragraph{\@startsection{paragraph}{4}{10pt}{-1.25ex plus -1ex minus -.1ex}{0ex plus 0ex}{\normalsize\textit}}
\renewcommand\@biblabel[1]{#1}
\renewcommand\@makefntext[1]%
{\noindent\makebox[0pt][r]{\@thefnmark\,}#1}
\makeatother
\renewcommand{\figurename}{\small{Fig.}~}
\sectionfont{\large}
\subsectionfont{\normalsize}

\fancyfoot{}
\fancyfoot[LO,RE]{\vspace{-7pt}\includegraphics[height=9pt]{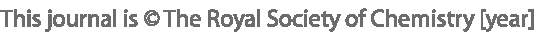}}
\fancyfoot[CO]{\vspace{-7.2pt}\hspace{12.2cm}\includegraphics{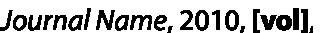}}
\fancyfoot[CE]{\vspace{-7.5pt}\hspace{-13.5cm}\includegraphics{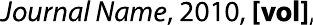}}
\fancyfoot[RO]{\footnotesize{\sffamily{1--\pageref{LastPage} ~\textbar  \hspace{2pt}\thepage}}}
\fancyfoot[LE]{\footnotesize{\sffamily{\thepage~\textbar\hspace{3.45cm} 1--\pageref{LastPage}}}}
\fancyhead{}
\renewcommand{\headrulewidth}{1pt}
\renewcommand{\footrulewidth}{1pt}
\setlength{\arrayrulewidth}{1pt}
\setlength{\columnsep}{6.5mm}
\setlength\bibsep{1pt}

\twocolumn[
  \begin{@twocolumnfalse}
\noindent\LARGE{\textbf{Quantum interference and electron correlation in charge
transport through triangular quantum dot molecules$^\dag$}}
\vspace{0.6cm}

\noindent\large{\textbf{Chih-Chieh Chen,\textit{$^{a}$} Yia-chung Chang,$^{\ast}$\textit{$^{abc}$} and
David M T Kuo$^{\ast}$\textit{$^{d}$}}}\vspace{0.5cm}

\noindent\textit{\small{\textbf{Received Xth XXXXXXXXXX 20XX, Accepted Xth XXXXXXXXX 20XX\newline
First published on the web Xth XXXXXXXXXX 200X}}}

\noindent \textbf{\small{DOI: 10.1039/b000000x}}
\vspace{0.6cm}

\noindent \normalsize{We study the charge transport properties
of triangular quantum dot molecule (TQDM) connected to
metallic electrodes, taking into account all correlation functions and
relevant charging states. The quantum interference (QI) effect of TQDM
resulting from electron coherent tunneling between quantum dots is revealed
and well interpreted by the long distance coherent tunneling mechanism. The spectra of
electrical conductance of TQDM with charge filling from one to six
electrons clearly depict the many-body and topological effects. The calculated
charge stability diagram for conductance and total occupation
numbers match well with the recent experimental measurements. We
also demonstrate that the destructive QI effect on the tunneling
current of TQDM is robust with respect to temperature variation, making the single electron
QI transistor feasible at higher temperatures.}
\vspace{0.5cm}
 \end{@twocolumnfalse}
  ]

\footnotetext{\dag~Electronic Supplementary Information (ESI) available. See DOI: 10.1039/b000000x/}


\footnotetext{\textit{$^{a}$~Department of Physics, University of Illinois at Urbana-Champaign, Urbana, Illinois 61801, USA.}}
\footnotetext{\textit{$^{b}$~Research Center for Applied Sciences, Academic Sinica, Taipei, 11529 Taiwan. E-mail: yiachang@gate.sinica.edu.tw}}
\footnotetext{\textit{$^{c}$~Department of Physics, National Cheng-Kung University, Tainan, 70101 Taiwan.}}
\footnotetext{\textit{$^{d}$~Department of Electrical Engineering and Department of Physics, National Central
University, Chungli, 320 Taiwan. E-mail: mtkuo@ee.ncu.edu.tw}}



\section{Introduction}

{Molecule transistors (MTs) provide a brightened
scenario of nanoelectronics with low power consumption.\cite{Reed1997,Joachim2000,Bergfield2009} To date,
the implementation of MTs remains challenging, and a good theoretical understanding of their characteristics is essential for advancing the technology. The
current-voltage (I-V) curves of MTs are typically predicted by calculations based on the
density functional theory (DFT).\cite{Bergfield2009} However, the DFT approach cannot fully capture the correlation effect in the transport behavior of MTs in the Coulomb-blockade regime.  A theoretical framework to treat adequately the many-body problem of a molecular junction remains elusive due
to the complicated quantum nature of such devices. Experimental studies of a
artificial molecule with simplified structures are important not only
for the advances of novel nanoelectronics, but also for providing a
testing ground of many body theory.} For example, the coherent
tunneling between serially coupled double quantum dots (DQDs) was
studied and demonstrated for application as a spin filter in the
Pauli spin blockade regime.\cite{Hanson2007} Recent experimental studies have
been extended to serially coupled triple quantum dots (SCTQDs) for
studying the effect of long distance coherent tunneling (LDCT) in
electron transport.\cite{Busl2013,Braakman2013,Amaha2013} Triangular quantum dot molecule (TQDM) provides the simplest topological structure with quantum interference (QI)
phenomena.\cite{Nitzan2003,Stafford2007,Poltl2013} The QI effect in the coherent tunneling process
of TQDM junctions has been studied experimentally.\cite{Seo2013,Guedon2012}

 It was suggested that the tunneling currents through benzene molecules
can also show a destructive QI behavior.\cite{Bergfield2009,Stafford2007} The tunneling current through a single benzene molecule was theoretically studied by DFT.\cite{Stafford2007} However,
the influence of the strong correlation on the QI effect remains unclear due to the limitation of DFT. Many theoretical works have pointed out that electron
Coulomb interactions have strong influence not only on the
electronic structures of TQDM,\cite{Korkusinski2007,Hsieh2012} but
also on the probability weights of electron transport
paths.\cite{Meir1992,Kuo2007,Bulka2004,Kuo2011} When both the intradot and interdot Coulomb
interactions in a TQDM are included, the transport behavior involving
multiple electrons becomes quite complicate. The setup for the TQDM junction of interest is depicted in Fig.~1. Here we present a full many-body solution to the
tunneling current of TQDM, which can well illustrate the Pauli spin
blockade effect of DQDs,$^{4}$ LDCT of SCTQDs$^{5-7}$ and QI of TQDMs$^{11}$
for both equilibrium and nonequilibrium cases. Thus, our theoretical work can provide
useful guidelines for the design of future molecular electronics and the realization of large scale quantum registers built by
multiple QDs.\cite{Loss1998}

\begin{figure}[t]
\centering
\includegraphics[scale=0.3]{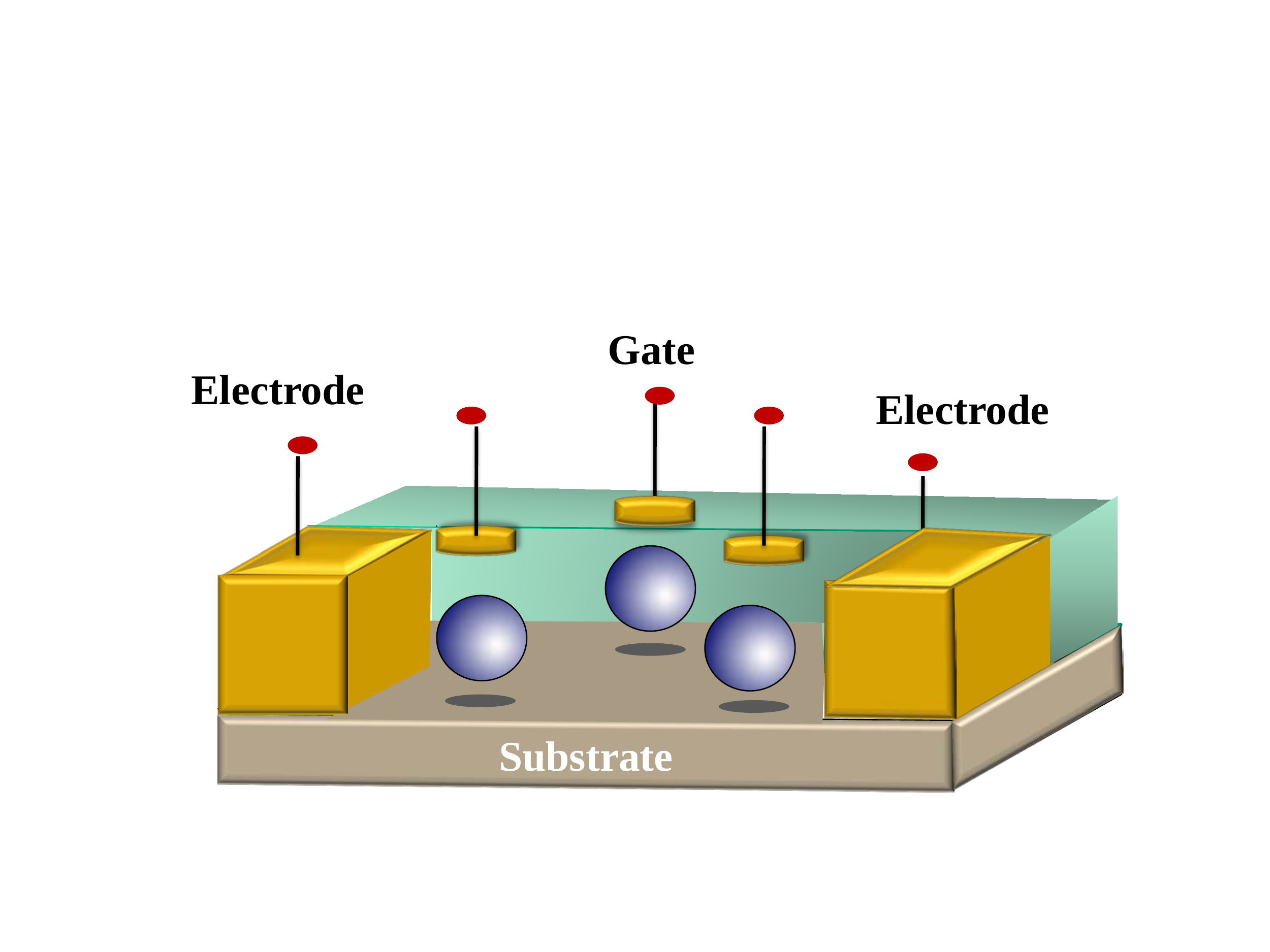}
\caption{Illustration of the TQDM junction system of interest.}
\end{figure}

We adopt the equation of motion method (EOM), which is a powerful tool for studying electron transport, taking into account electron Coulomb interactions.\cite{Meir1992,Kuo2007,Bulka2004,Kuo2011} This method has been
applied to reveal the transport behaviors of a single QD with
multiple energy levels\cite{Kuo2007} and DQDs.\cite{Bulka2004,Kuo2011} For a TQDM
with one level per QD, there are 64 configurations for electrons to
transport between electrodes.\cite{Weymann2011} Previous theoretical works have ignored
the high-order Green functions resulting from electron Coulomb
interactions to simplify the calculation.\cite{Weymann2011}  To have a full
solution becomes crucial for depicting the charge transport involving a few
electrons. We solve the EOM of Green functions up to six electrons, taking into account
all correlations caused by electron Coulomb interactions
and electron hopping between TQDM. This involves solving 4752 Green's
functions and 923 correlation functions self-consistently.

\section{Model}

We consider an artificial molecule made of nanoscale QDs, in which the energy
level separations are much larger than the on-site Coulomb
interactions and thermal energies. Thus, only one energy level for
each quantum dot is included. The extended
Hubbard-Anderson model is employed to simulate the TQDM junction
with Hamiltonian given by {$H=H_0+H_T+H_{QDs}$, where
$H_0=\sum_{{\bf k},\sigma,\alpha}\epsilon_{{\bf k}}
c^{\dagger}_{{\bf k},\sigma,\alpha}c_{{\bf k},\sigma,\alpha}$ is the Hamiltonian
for free electrons in the electrodes.
$c^{\dagger}_{k,\sigma,\alpha}$($c_{{\bf k},\sigma,\alpha}$) creates
(destroys) an electron of momentum {\bf k} and spin $\sigma$ with energy
$\epsilon_k$ in the $\alpha$ electrode. $H_T=\sum_{{\bf k},\ell,\alpha}
(V_{k,\alpha,\ell} c^{\dagger}_{{\bf k},\sigma,\alpha}
d_{\ell,\sigma}+V^*_{{\bf k},\alpha,\ell} d^{\dagger}_{\ell,\sigma}
c_{{\bf k},\sigma,\alpha})$. $V_{{\bf k},\alpha,\ell}$ describes the coupling
between the $\alpha$ electrode and the $\ell$-th QD.
$d^{\dagger}_{\ell,\sigma}$ ($d_{\ell,\sigma}$) creates (destroys)
an electron in the $\ell$-th dot.} $H_{QDs}$ is the
extended Hubbard Hamiltonian for multiple QDs.
\begin{eqnarray}
H_{QDs}&=& \sum_{\ell,\sigma} E_{\ell} n_{\ell,\sigma}+
\sum_{\ell} U_{\ell} n_{\ell,\sigma} n_{\ell,\bar\sigma}
\\
&&+\sum_{\ell<j,\sigma,\sigma'} U_{\ell
j}n_{\ell,\sigma}n_{j,\sigma'} +\sum_{\ell\ne j,\sigma}t_{\ell j}
d^{\dagger}_{\ell,\sigma} d_{j,\sigma} ,\notag
\end{eqnarray}
where $E_{\ell}$ is the spin-independent QD energy level,
$n_{\ell,\sigma}=d^{\dagger}_{\ell,\sigma}d_{\ell,\sigma}$,
$U_{\ell}$ and $U_{\ell j}$ ($\ell< j$) denote the intradot and
interdot Coulomb interactions, respectively and $t_{\ell j}$
describes the electron interdot coupling. {The interdot
Coulomb interactions as well as intradot Coulomb interactions are
important for nanoscale semiconductor QDs and molecules. Therefore,
$U_{\ell,j}$ cannot be ignored.}

Using the Keldysh-Green's function technique\cite{Meir1992,Haug2007}, the electrical
current from reservoir $\alpha$ to the TQDM junction is calculated
according to the Meir-Wingreen formula
\begin{equation}
J_\alpha =\frac{ie}{h}\int {d\epsilon} \sum_{j\sigma}
\Gamma^\alpha_{j} [ G^{<}_{j\sigma} (\epsilon) + f_\alpha
(\epsilon)(  G^{r}_{j\sigma}(\epsilon) -
G^{a}_{j\sigma}(\epsilon) ) ],
\end{equation}
where {$\Gamma^\alpha_j(\epsilon)=\sum_{\bf k}
|V_{{\bf k},\alpha,j}|^2 \delta(\epsilon-\epsilon_{\bf k})$  is the tunneling rate between the
$\alpha$-th reservoir and the $j$-th QD. Throughout the paper, for two-terminal devices we assume that the left (right) lead is only coupled to the left (right) QD with tunneling rate $\Gamma_L$ ($\Gamma_R$), while there is no coupling between the center QD and the two leads. For three-terminal devices, the coupling between the center QD and a third gate is described by the tunneling rate $\Gamma_C$.}
$f_{\alpha}(\epsilon)=1/\{\exp[(\epsilon-\mu_{\alpha})/k_BT]+1\}$
denotes the Fermi distribution function for the $\alpha$-th
electrode, where $\mu_\alpha$ is the chemical potential and $T$ is
the temperature of the system. $e$, $h$, and $k_B$ denote the
electron charge, the Planck's constant, and the Boltzmann constant,
respectively. $G^{<}_{j\sigma} (\epsilon)$,
$G^{r}_{j\sigma}(\epsilon)$, and $G^{a}_{j\sigma}(\epsilon)$ are the
frequency domain representations of the one-particle lessor,
retarded, and advanced Green's functions
$G^{<}_{j\sigma}(t,t')=i\langle d_{j,\sigma}^\dagger (t')
d_{j,\sigma}(t) \rangle $, $G^{r}_{j\sigma}(t,t')=-i\theta
(t-t')\langle \{ d_{j,\sigma}(t),d_{j,\sigma}^\dagger (t') \}
\rangle $, and $G^{a}_{j\sigma}(t,t')=i\theta (t'-t)\langle \{
d_{j,\sigma}(t),d_{j,\sigma}^\dagger (t') \}  \rangle $,
respectively. {These one-particle Green's functions are related
recursively to other Green's functions and correlators via the
many-body equation of motion,\cite{Kuo2007,Bulka2004,Kuo2011} which we solve via an
iterative numerical procedure to obtain all $n$-particle Green's
functions ($n=1,\cdots, 6$) and correlators for the TQDM. (See
supplemental materials.) Our procedure is valid in the Coulomb
blockade regime, but not the Kondo regime.\cite{Goldhaber1998,Numata2009} Throughout this
paper, we assume the on-site Coulomb interaction
$U_{\ell}=U_0=100\Gamma_0$ for all three QDs and the same tunneling
rates at all leads, $\Gamma^{\alpha}_{j}=\Gamma$ with $j$ labeling
the QD directly connected to lead $\alpha$.}

\section{Results and discussion}

\begin{figure}[t]
\centering
\includegraphics[scale=0.3]{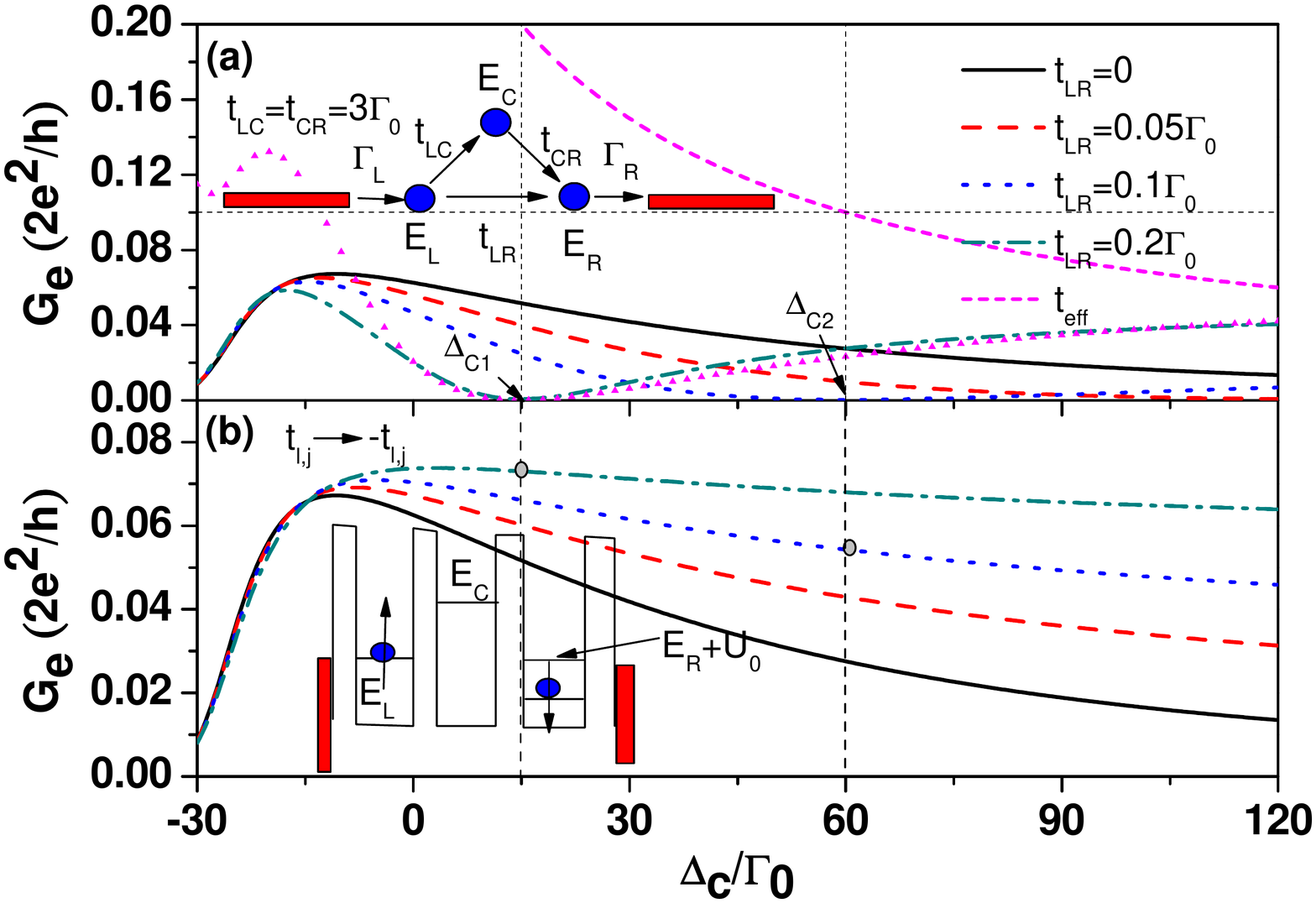}
\caption{Electrical conductance ($G_e$) of TQDM as a function of
central QD energy for different $t_{LR}$ strengths in the Pauli spin
blockade configuration with $E_L=E_F$ and $E_R=E_F-U_R$. (a)
$t_{LC}=t_{CR}=3\Gamma_0$, and (b) $t_{\ell,j}$ replaced by
$-t_{\ell,j}$. Other physical parameters are $k_BT=1\Gamma_0$,
$\Gamma=0.3\Gamma_0$, $U_{LC}=U_{CR}=30\Gamma_0$, and $U_{LR}=0$.}
\end{figure}

{ Although the QI of TQDM was theoretically investigated previously\cite{Michaelis2006,Emary2007}, the effect of intradot and interdot Coulomb interactions was not considered. Here, we utilize the LDCT effect to tune the effective hopping strength between outer QDs to achieve the
destructive and constructive QIs in the presence of electron Coulomb
interaction.} We consider a TQDM junction in the Pauli spin blockade
(PSB) configuration\cite{Hanson2007,Kuo2011} with $E_L=E_F$, $E_R=E_F-U_{0}$,
$t_{LC}=t_{CR}=3\Gamma_0$, $\Gamma_L=\Gamma_R=0.3\Gamma_0$,
$U_{LC}=U_{CR}=30\Gamma_0$, and $U_{LR}=0$. Fig.~2(a) shows the
electrical conductance ($G_e$) as a function of central QD energy
level ($\Delta_C=E_C-E_F$) for $t_{LR}$ varying from 0 to
$0.2\Gamma_0$ at $k_BT=1\Gamma_0$ (in weak interdot coupling
regime). For $\Delta_C$ less than $-15\Gamma_0$, $G_e$ is not
sensitive to the variation of $t_{LR}$, indicating that the
transport is mainly through the upper path involving the center QD
as shown in the inset of Fig.~2(a). In the case of $t_{LR}=0$, $G_e$
can be well explained by the LDCT effect when $E_C$ is far away from
$E_F$.\cite{Kuo2014} The central QD provides an intermediated state for
electrons in the outer QDs. Through the upper path, TQDM behaves
like a double QD with an effective hopping
$t_{eff}=-t_{LC}t_{CR}/(U_{CR}+\Delta_C)$, which can also be
understood by the second order perturbation theory.\cite{Kuo2014} Once
$t_{LR} \neq 0$  (the lower path turns on), electron transport
through the two paths with $t_{eff}$ and $t_{LR}$ lead to a
destructive QI. Note that $G_e$ for the case of $t_{LR} \neq 0$ is
reduced compared to the case of $t_{LR}=0$. In particular, $G_e$ is
vanishingly small at the value of $\Delta_C$ where $|t_{eff}|$
crosses $t_{LR}$ as indicated by dashed lines in Fig.~2(a). For
illustration, the curve for
$|t_{eff}|=|t_{LC}t_{CR}|/(U_{CR}+\Delta_C)$  is also shown in
Fig.~2(a). (See short-dashed curve) The vanishing $G_e$ occurs at
lower $\Delta_C$ with increasing $|t_{LR}|$ (Compare dash-dotted
with dashed curves). Due to topological effect, the electron-hole
symmetry does not hold for the energy spectrum of TQDM.\cite{Korkusinski2007} {When TQDM has identical QD energy levels ($E_{\ell}=E_0$) and homogenous
electron hopping strengths $t_{\ell,j}=t$, we have one level
$\epsilon=E_0+2t$ and one doubly degenerate level with
$\epsilon=E_0-t$ for the case of $U_{\ell}=U_{\ell,j}=0$. Unlike the
cases of DQDs and SCTQDs, the lowest energy level depends on the
sign of $t_{\ell,j}$. This is a manifesting result of electron-hole
asymmetrical behavior of TQDM. Therefore, it will be interesting to examine the sign effect of
$t_{\ell,j}$ on the QI behavior. Physically, the sign of $t_{\ell,j}$ depends on the symmetry properties of orbitals, which can change in different configurations.} We can
replace $t_{\ell,j}$ by $-t_{\ell,j}$ to examine the QI effect with
respect to the electron-hole symmetry. The results are shown in
Fig.~2(b). We find that $G_e$ is enhanced with increasing $t_{LR}$,
which is attributed to the constructive QI effect, in contrast to
the destructive QI effect shown in Fig.~2(a).

To gain deeper understanding of the destructive and constructive QI
shown in Fig.~2, we compare our full calculation with the weak
interdot-coupling theory,\cite{Kuo2011,Kuo2014} which allows simple closed-form
expression for the electrical conductance of TQDM. We obtain $G_e =
2e^2/h \int d\epsilon {\cal T}(\epsilon) [\partial
f(\epsilon)/\partial E_F]\approx (2e^2/h) {\cal T}(E_F)$ at
low-temperature limit, where the transmission coefficient ${\cal
T}(\epsilon)$ with 64 configurations is approximately given by
\begin{eqnarray}
&&{\cal T}_{PSB}(\epsilon) \label{PSB}  \\
&&=\frac{4\Gamma_L\Gamma_R P_{PSB}F_{QI}}
{|\mu_1\mu_2\mu_3-t^2_{CR}\mu_1-t^2_{LC}\mu_3-t^2_{LR}\mu_2-2t_{LR}t_{LC}t_{CR}|^2}, \notag
\end{eqnarray}
where $F_{QI}=\mu^2_2(t_{LC}t_{CR}/\mu_2+t_{LR})^2$ is a factor
related to QI. $\mu_1=\epsilon-E_L+i\Gamma_L$,
{$\mu_2=\epsilon-E_C-U_{RC}$} and
$\mu_3=\epsilon-E_R-U_R+i\Gamma_R$. $P_{PSB}$ denotes the
probability weight in the PSB configuration.\cite{Kuo2011} From Eq.~(3), we
have

\begin{equation}
G_e=\frac{2e^2}{h}
\frac{P_{PSB}4\Gamma^2(t_{eff}+t_{LR})^2}{(\Gamma^2+2t_{eff}t_{LR}+t^2_{LR})^2+\Gamma^2~(t_{h1}+t_{h2})^2},
\label{PSB1}
\end{equation}
where $t_{eff}=-t_{LC}t_{CR}/(U_{CR}+\Delta_C)$,
$t_{h1}=-t^2_{LC}/(U_{CR}+\Delta_C)$, and
$t_{h2}=-t^2_{CR}/(U_{CR}+\Delta_C)$ {with $\Gamma_L=\Gamma_R\equiv \Gamma$}. For $t_{LC}=t_{CR}=t_c=0$, Eq.~(4) reduces to the conductance of DQD,$^{17}$ while for $t_{LR}=0$, it reduces to the $G_e$ of SCTQD.\cite{Kuo2014} At $\Delta_C=15\Gamma_0$ and
$60\Gamma_0$, which satisfy the condition of $t_{eff}+t_{LR}=0$ for
$t_{LR}=0.2\Gamma_0$ and $t_{LR}=0.1\Gamma_0$, respectively, and we
see $G_e$ vanishes there. This well illustrates the destructive QI
seen in Fig.~2(a). Once we make the substitution
$t_{\ell,j}\rightarrow -t_{\ell,j}$ in Eq. (4), we can reveal the
constructive QI in $G_e$ as shown in Fig.~2(b).  Note that in the
weak coupling regime ($t_{eff}/\Gamma \ll 1$), the probability
weight $P_{PSB}$ of Eq.~(4) calculated according to the procedures
in Ref.~\citenum{Kuo2011}, where the interdot two-particle correlation functions are
factorized as the product of single occupation numbers, is
consistent with the full calculation, but not for $t_{eff}/\Gamma
\gg 1$. Away from the weak coupling regime, the interdot electron
correlations become important. {To explicitly reveal the importance of electron
correlation effects, we plot the curve of $t_{LR}=0.2\Gamma_0$ (with
triangle marks) calculated by the procedure of Ref. 18 (including 64
configurations) in Fig.~2(a). Comparison between the full solution
and the approximation considered in Ref. 18, we find that the
electron correlation effects become very crucial when
$t_{eff}/\Gamma \gg 1$. Once electron transport involves more
electrons, the high-order (beyond two-particle) Green functions and correlation functions
should be included (see the results of Fig. 3).
The difference between the conventional mean-field theory of Ref.~3 with the full solution is even larger.
The comparison between mean-field theory and the procedure of Ref.~18 has been discussed in the appendix of Ref.~18.}

According to Eq.~(4), constructive and destructive QI
effects depend on the sign of $t_{\ell,j}$. Therefore, if the
wavefunction of the center dot has opposite parity (say, an $x$-like
state) with respect to the wavefunctions in two outer dots (assumed
to be $s$-like), then $t_{LC}$ will have opposite sign compared with
$t_{CR}$ and $t_{LR}$, and the sign of $t_{eff}$ will be flipped.
Consequently, the destructive QI shown in Fig.~2(a) will become
constructive QI. Thus, for a center QD with an $s$-like ground state
and an excited $p$-like state, it is possible to see the change of
QI between destructive and constructive by tuning the gate voltage,
which sweeps through different resonance energies of the center QD
in addition to the change of sign of $t_{eff}$ when the Fermi level
goes from below the resonance level to above the resonance level.
From the results of Fig.~2, QI effect can be electrically controlled
by the energy level $E_c$. This advantage of TQDM may be useful for
improving the spin filtering of DQDs \cite{Hanson2007}.

\begin{figure}[t]
\centering
\includegraphics[scale=0.3]{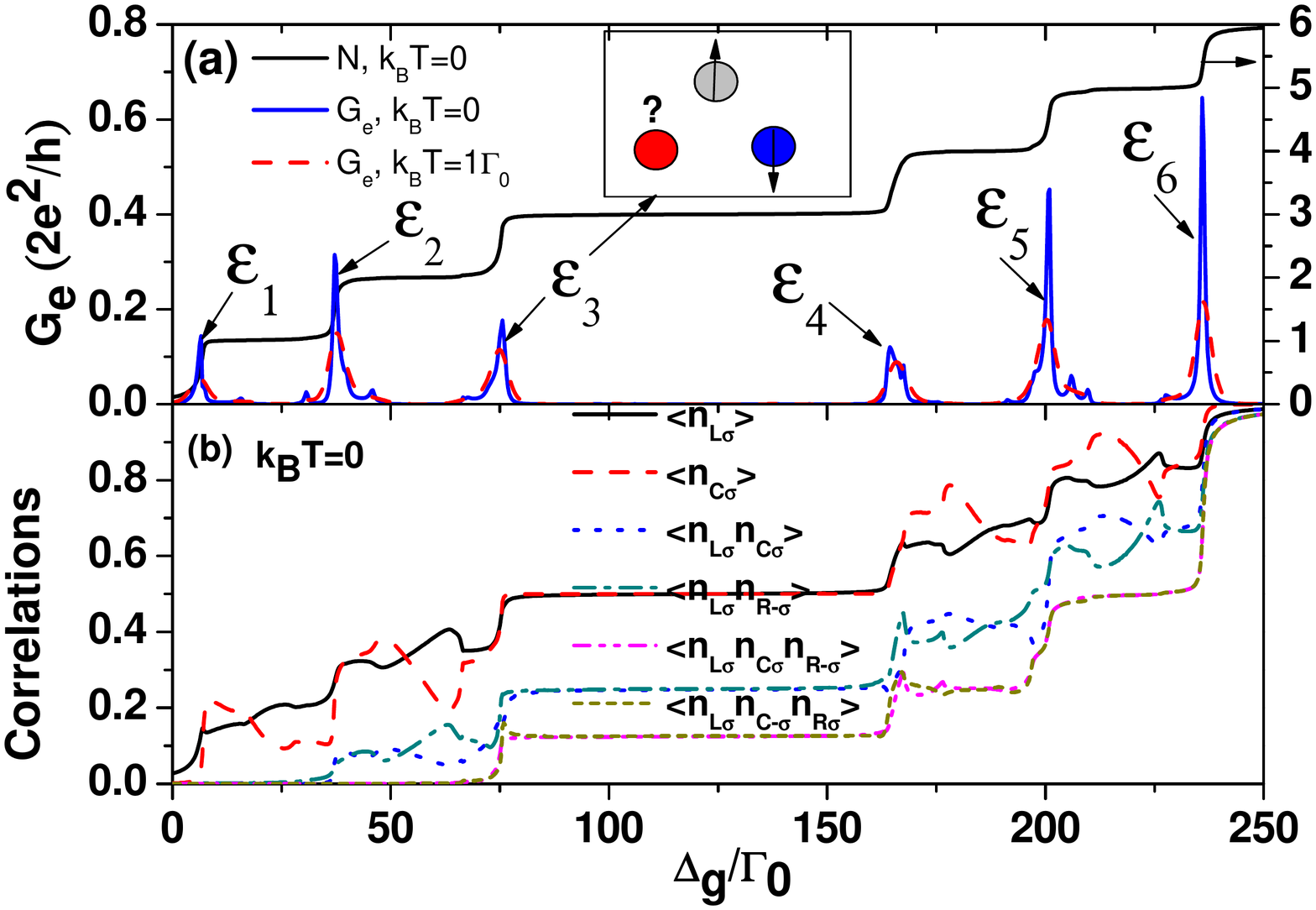}
\caption {(a) Electrical conductance of TQDMs as a
function of gate voltage { $\Delta_g$ with $E_L=E_R=E_C=E_F+10\Gamma_0-\Delta_g$
and} $\Gamma=1.6\Gamma_0$.(b) Correlation functions for
$k_BT=0\Gamma_0$. Other physical parameters are
$t_{\ell,j}=3\Gamma_0$ and $U_{LR}=U_{LC}=U_{CR}=30\Gamma_0$.}
\end{figure}

In addition to QI effects, spin frustration and topological effects
(due to electron-hole asymmetry) on the measured quantities
(electrical conductance or current) are {also interesting
issues.\cite{Seo2013} Fig.~3(a) shows $G_e$ as a function of gate voltage,
$\Delta_g$ for $E_{\ell}=E_F+10\Gamma_0-\Delta_g$; $\ell=L,C,R$ at
two different temperatures, $k_BT=0$ and $1\Gamma_0$. Here, we
consider the homogenous configuration with
$U_{LR}=U_{LC}=U_{CR}=30\Gamma_0$ and $t_{\ell,j}=t_c=3\Gamma_0$. At
low temperature, there are six main peaks in the $G_e$ spectrum,
labeled by $\epsilon_n$, $n=1,\cdots,6$ and some secondary peaks. At
higher temperature, the six main peaks are suppressed and broadened
as shown by the dashed curve, and the secondary peaks are washed
out. We also plot the total occupation number,
$N=\sum_{\sigma}(N_{L,\sigma}+N_{R,\sigma}+N_{C,\sigma})$ as the
black solid curve, which shows a stair-case behavior with plateaus
at $N=1,\cdots, 6$, corresponding to the filling of TQDM with 1 to
6 electrons. It can be seen that the six main peaks occur at
$\Delta_g$ where $N$ is increased by 1. Thus, the peak positions
$\epsilon_n$ correspond to the chemical potential of electrons in
TQDM, i.e. the energy needed to add an electron to the system. The
main peak positions can be approximately obtained by the calculation
of chemical potential of TQDM without considering the coupling with
leads as done in Ref.~\citenum{Korkusinski2007}.  For example,
$\epsilon_{1}=E_L-2|t_c|$,
$\epsilon_2=E_L+U_{LC}-8t^2/(U_{0}-U_{LC})$ and
$\epsilon_{3}=E_L+2U_{\ell,j}-3J_{ex}/2+2|t_c|+16t^2_c/(U_{0}-U_{LC})$
under the condition $U_{0} > U_{LC} \gg t_{c}$, where $J_{ex}\equiv
E_{0}(S=3/2)-E_{0}(S=1/2)$ is the difference in energy between the
spin-3/2 and spin-1/2 configuration.\cite{Korkusinski2007} However, the relative strengths of
peaks in the conductance spectrum can only be obtained by solving
the full Anderson-Hubbard model self-consistently.

Unlike the $G_e$ spectrum of DQDs,\cite{Bulka2004} the $G_e$ spectrum of
TQDM does not show the electron-hole symmetry due to topological
effect. Note that $N=4$ and $N=5$ correspond to two-hole and
one-hole configurations, respectively. A large Coulomb blockade
separation between $\epsilon_3$ and $\epsilon_4$ is given by
$\Delta_{34}=U_0+3J_{ex}-4t_c-8t^2_c/(U_0-U_{LC})$. Here,
$\epsilon_4$ corresponds to the two hole ground state with spin
triplet instead of singlet.  The magnitude of $G_e$ is smaller than
the quantum conductance $2e^2/h$ for $t_{LR}/\Gamma \gg 1$ as a
result of electron Coulomb interactions.\cite{Bulka2004} The mechanism for
understanding the unusual $G_e$ behavior in nanostructure junction
systems is a subject of high interest.\cite{Bauer2013} Due to electron
Coulomb interactions, the magnitudes of peaks are related to the
probability weights of quantum paths, which are related to
single-particle occupation numbers and many-particle correlation
functions.\cite{Kuo2011}}

To reveal the configurations for each main peak, the one-particle
occupation number $N_{\ell,\sigma}\equiv \langle
n_{\ell,\sigma}\rangle$, interdot two particle correlation functions
$\langle n_{\ell,\sigma} n_{j,\bar\sigma} \rangle$ , and three
particle correlation functions ($\langle n_{L,\sigma}n_{C,-\sigma}
n_{R,\sigma}\rangle$, and $\langle n_{L,\sigma}n_{C,-\sigma}
n_{R,\sigma}\rangle$) are plotted in Fig.~3(b). We always have the
relation $N_{L,\sigma}=N_{R,\sigma}\neq~N_{C,\sigma}$, because the
outer QDs are directly coupled to electrodes, {but not the central
QD. Such a relation also holds for two-particle and three-particle
correlation functions. The six main peaks in Fig.~3(a) indicate the
filling of TQDM up to the $n$-electron ground state for
$n=1,\cdots,6$. For example, $\epsilon_2$ indicates the formation of
two-electron state with spin singlet, while $\epsilon_3$ indicates
the formation of three-particle state with total spin $S=1/2$, which
can be described as the spin-frustration state\cite{Seo2013,Korkusinski2007,Andergassen2013}. Because
the on-site Coulomb interaction favors homogeneous distribution of
three electrons in TQDM, whereas the interdot Coulomb repulsion
favors the charge fluctuation. As seen in Fig.~3(b), for $\Delta_g
\le \epsilon_3$ ($~78 \Gamma_0$) $N_{\ell}$ in each dot clearly
displays the charge fluctuation behavior. When TQDM goes into a
three-particle state ($\Delta_g > \epsilon_3$), the charge
fluctuation is suppressed, and each QD is filled with one particle
(with $N_{L,\sigma}=N_{R,\sigma}=N_{C,\sigma}=0.5$), while $\langle
n_{L,\sigma}n_{C,\sigma} n_{R,-\sigma}\rangle$=$\langle
n_{L,\sigma}n_{C,-\sigma} n_{R,\sigma}\rangle =\langle
n_{L,-\sigma}n_{C,\sigma} n_{R,-\sigma}\rangle$. This also
demonstrates the spin frustration condition as depicted in the inset
of Fig.~3(a).}

\begin{figure}[t]
\centering
\includegraphics[scale=0.5,trim=200 200 200 200]{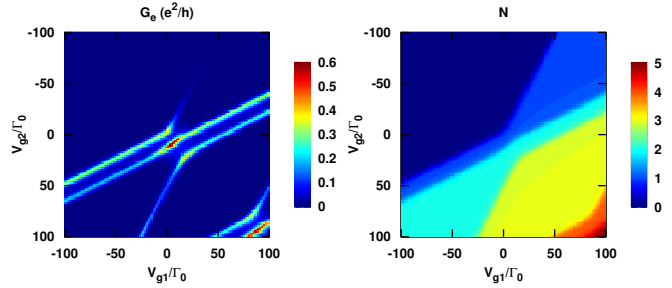}
\caption{Charge stability diagram of TQDM. $\Gamma_\ell=\Gamma_0$,
$k_BT=1.5\Gamma_0$, $U_{LC}=U_{CR}=U_{LR}=12\Gamma_0$,
$t_{\ell,j}=3\Gamma_0$. The energy levels are shifted according to
$E_\ell=E_F-U_{LR}-\sum_{m=1}^2 \beta_{\ell,m}eV_{gm}$, where the
gate coupling constants are $\beta_{L,1}=0.5$, $\beta_{C,1}=1$,
$\beta_{R,1}=0.5$, $\beta_{L,2}=1$, $\beta_{C,2}=0.5$, and
$\beta_{R,2}=1$.}
\end{figure}

{Figure~4 shows the charge stability diagram for zero-bias
electrical conductance ($G_e$) and total occupation number ($N$) as
functions of gate voltages exerted on any two QDs (labeled by
$V_{g1}$ and $V_{g2}$) for a TQDM connected to three terminals.
 The magnitudes
of $G_e$ and $N$ are indicated by different colors. It is noticed
that $G_e$ is enhanced on the borders that separate domains of
different values of occupation number ($N$) with larger $G_e$
occurring at $V_{g1}=V_{g2}$. This is a result of higher degeneracy
and charge-fluctuation in the state. The largest $G_e$ for $N\le 3$
occurs at the junction between $N=1$ and $N=2$ domains when
$V_{g1}=V_{g2}$. This feature corresponds to the $\epsilon_2$ peak
of Fig.~3(a). The diagram Fig.~4(a) is simply a collection of curves
displayed in Fig.~3(a) at different values of $V_{g2}$ that shifts
the QD energy levels. We note that in the domains of $N=1$ and
$N=2$, the areas with stripes are not symmetrical with respect to
gate voltage. In Ref.~\citenum{Seo2013}, a capacitive interaction model was
employed to plot the diagram of $N$. In their model, the electron
hopping strength $t_{\ell,j}$ was ignored. Consequently, the charge
stability diagram of $G_e$ cannot be obtained. The charge stability
diagram of $G_e$ obtained by our full calculation [as shown in Fig.~4(a)] bears close resemblance to the experimental results as shown
in Fig.~2 of Ref.~\citenum{Seo2013}.

\begin{figure}[t]
\centering
\includegraphics[scale=0.3]{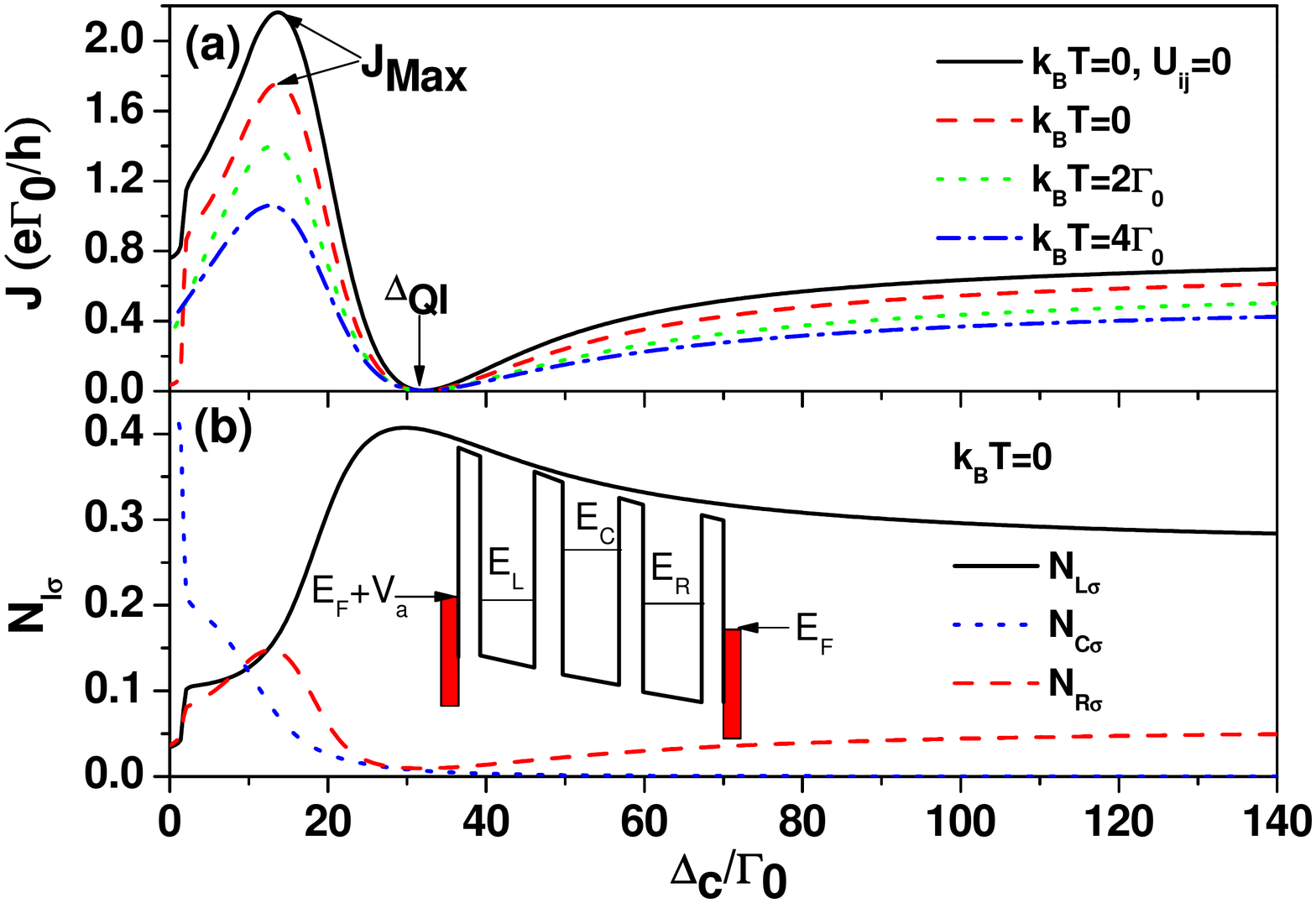}
\caption{(a) Tunneling current as a function of $\Delta_C=E_C-E_F$
at finite bias $V_a=10\Gamma_0$ for various temperatures. (b)
Occupation numbers as a function of $\Delta_C$ at $k_BT=0$. The
other physical parameters used are $E_L=E_R=E_F+10\Gamma_0$,
$U_{LC}=U_{CR}=30\Gamma_0$, $U_{LR}=10\Gamma_0$,
$t_{LC}=t_{CR}=3\Gamma_0$, $t_{LR}=0.4\Gamma_0$, and
$\Gamma=\Gamma_0$.}
\end{figure}

So far, the results shown in Figs.~2-4 are all related to the linear
response. To further clarify the QI effect at finite bias
($V_a=10\Gamma_0$) for different temperatures, we plot in Fig.~5(a)
the tunneling current as a function of center QD energy,
$E_C=E_F+\Delta_c$ for the configuration shown in the inset of
Fig.~5(b). For $\Delta_C\ge 2\Gamma_0$, the tunneling current is
suppressed as temperature increases. This is attributed to a
reduction of electron population in the electrodes for electrons
with energy near $E_F+10\Gamma_0$. For $\Delta_C\le 2\Gamma_0$, we
notice that $N_{c,\sigma}$ quickly jumps to 0.4, indicating that the
central QD is filled with charge, which causes an interdot Coulomb
blockade for  electrons entering the left QD. (See the reduction of
$N_{L,\sigma}$ in Fig.~5(b)). This explains the sharp dip of $G_e$
for $\Delta_C\le 2 \Gamma_0$ in Fig.~5(a). As temperature increases,
such a dip in tunneling current is smeared out. At
$\Delta_{C}=10\Gamma_0+t_{LC}t_{CR}/t_{LR}\equiv \Delta_{QI}$, the
tunneling current vanishes for all temperatures considered due to
the QI effect. Such a robust destructive QI effect with respect to
temperature provides a remarkable advantage for the realization of
single electron QI transistors at room temperature.\cite{Guedon2012} To
understand the interdot correlation effect, we also plot the case
without interdot Coulomb interaction ($U_{\ell,j}=0$) for $k_BT=0$
in Fig.~5(a). (See solid curve) We notice that the QI effect remains
qualitatively the same, except that the tunneling current is
slightly enhanced with interdot Coulomb interaction turned off. As
the QI effect suppresses the current flow, the charge will
accumulate in the left dot. Thus, $N_{L,\sigma}$ reaches the maximum
at $\Delta_C = \Delta_{QI}$ while $N_R$ reaches the minimum as seen
in Fig.~5(b). This implies that the QI effect can be utilized to
control charge storage in TQDM.

\section{Conclusions}
In summary we have obtained full solution to the charge
transport through TQDM junction in the presence of electron Coulomb
interactions, which includes all $n$-electron ($n=1,\cdots,6$)
Green's functions and correlation functions. The destructive and
constructive QI behaviors of TQDM are clarified {by considering the LDCT effect on the conductance spectrum. The conductance spectrum of
TQDM with total occupation number varying from one to six directly
reveals the electron-hole asymmetry due to topological effect. The
calculated correlation functions also illustrate the charge
fluctuation and spin frustration behaviors of TQDM.} Our
numerical results for charge stability diagram match experimental
measurements very well. Finally, we demonstrated that the QI effect
in TQDM is robust against temperature variation and it can be
utilized to control the charge storage.}

\mbox{}\\
{\bf Acknowledgments}\\
This work was supported in part by the National Science Council, Taiwan
under Contract Nos. NSC 101-2112-M-001-024-MY3 and NSC 103-2112-M-008-009-MY3.

\footnotesize{
\bibliography{TQDM_RSC} 

\providecommand*{\mcitethebibliography}{\thebibliography}
\csname @ifundefined\endcsname{endmcitethebibliography}
{\let\endmcitethebibliography\endthebibliography}{}
\begin{mcitethebibliography}{28}
\providecommand*{\natexlab}[1]{#1}
\providecommand*{\mciteSetBstSublistMode}[1]{}
\providecommand*{\mciteSetBstMaxWidthForm}[2]{}
\providecommand*{\mciteBstWouldAddEndPuncttrue}
  {\def\EndOfBibitem{\unskip.}}
\providecommand*{\mciteBstWouldAddEndPunctfalse}
  {\let\EndOfBibitem\relax}
\providecommand*{\mciteSetBstMidEndSepPunct}[3]{}
\providecommand*{\mciteSetBstSublistLabelBeginEnd}[3]{}
\providecommand*{\EndOfBibitem}{}
\mciteSetBstSublistMode{f}
\mciteSetBstMaxWidthForm{subitem}
{(\emph{\alph{mcitesubitemcount}})}
\mciteSetBstSublistLabelBeginEnd{\mcitemaxwidthsubitemform\space}
{\relax}{\relax}

\bibitem[Reed \emph{et~al.}(1997)Reed, Zhou, Muller, Burgin, and
  Tour]{Reed1997}
M.~A. Reed, C.~Zhou, C.~J. Muller, T.~P. Burgin and J.~M. Tour, \emph{Science},
  1997, \textbf{278}, 252--254\relax
\mciteBstWouldAddEndPuncttrue
\mciteSetBstMidEndSepPunct{\mcitedefaultmidpunct}
{\mcitedefaultendpunct}{\mcitedefaultseppunct}\relax
\EndOfBibitem
\bibitem[Joachim \emph{et~al.}(2000)Joachim, Gimzewski, and
  Aviram]{Joachim2000}
C.~Joachim, J.~Gimzewski and A.~Aviram, \emph{Nature}, 2000, \textbf{408},
  541--548\relax
\mciteBstWouldAddEndPuncttrue
\mciteSetBstMidEndSepPunct{\mcitedefaultmidpunct}
{\mcitedefaultendpunct}{\mcitedefaultseppunct}\relax
\EndOfBibitem
\bibitem[Bergfield and Stafford(2009)]{Bergfield2009}
J.~Bergfield and C.~Stafford, \emph{Phys. Rev. B}, 2009, \textbf{79},
  245125\relax
\mciteBstWouldAddEndPuncttrue
\mciteSetBstMidEndSepPunct{\mcitedefaultmidpunct}
{\mcitedefaultendpunct}{\mcitedefaultseppunct}\relax
\EndOfBibitem
\bibitem[Hanson \emph{et~al.}(2007)Hanson, Kouwenhoven, Petta, Tarucha, and
  Vandersypen]{Hanson2007}
R.~Hanson, L.~P. Kouwenhoven, J.~R. Petta, S.~Tarucha and L.~M.~K. Vandersypen,
  \emph{Rev. Mod. Phys.}, 2007, \textbf{79}, 1217--1265\relax
\mciteBstWouldAddEndPuncttrue
\mciteSetBstMidEndSepPunct{\mcitedefaultmidpunct}
{\mcitedefaultendpunct}{\mcitedefaultseppunct}\relax
\EndOfBibitem
\bibitem[Busl \emph{et~al.}(2013)Busl, Granger, Gaudreau, S{\'a}nchez, Kam,
  Pioro-Ladri{\`e}re, Studenikin, Zawadzki, Wasilewski, Sachrajda, and
  Platero]{Busl2013}
M.~Busl, G.~Granger, L.~Gaudreau, R.~S{\'a}nchez, A.~Kam,
  M.~Pioro-Ladri{\`e}re, S.~A. Studenikin, P.~Zawadzki, Z.~R. Wasilewski, A.~S.
  Sachrajda and G.~Platero, \emph{Nature nanotechnology}, 2013, \textbf{8},
  261--265\relax
\mciteBstWouldAddEndPuncttrue
\mciteSetBstMidEndSepPunct{\mcitedefaultmidpunct}
{\mcitedefaultendpunct}{\mcitedefaultseppunct}\relax
\EndOfBibitem
\bibitem[Braakman \emph{et~al.}(2013)Braakman, Barthelemy, Reichl, Wegscheider,
  and Vandersypen]{Braakman2013}
F.~R. Braakman, P.~Barthelemy, C.~Reichl, W.~Wegscheider and L.~M.~K.
  Vandersypen, \emph{Nature nanotechnology}, 2013, \textbf{8}, 432--437\relax
\mciteBstWouldAddEndPuncttrue
\mciteSetBstMidEndSepPunct{\mcitedefaultmidpunct}
{\mcitedefaultendpunct}{\mcitedefaultseppunct}\relax
\EndOfBibitem
\bibitem[Amaha \emph{et~al.}(2013)Amaha, Izumida, Hatano, Teraoka, Tarucha,
  Gupta, and Austing]{Amaha2013}
S.~Amaha, W.~Izumida, T.~Hatano, S.~Teraoka, S.~Tarucha, J.~A. Gupta and D.~G.
  Austing, \emph{Phys. Rev. Lett.}, 2013, \textbf{110}, 016803\relax
\mciteBstWouldAddEndPuncttrue
\mciteSetBstMidEndSepPunct{\mcitedefaultmidpunct}
{\mcitedefaultendpunct}{\mcitedefaultseppunct}\relax
\EndOfBibitem
\bibitem[Nitzan and Ratner(2003)]{Nitzan2003}
A.~Nitzan and M.~A. Ratner, \emph{Science}, 2003, \textbf{300},
  1384--1389\relax
\mciteBstWouldAddEndPuncttrue
\mciteSetBstMidEndSepPunct{\mcitedefaultmidpunct}
{\mcitedefaultendpunct}{\mcitedefaultseppunct}\relax
\EndOfBibitem
\bibitem[Stafford \emph{et~al.}(2007)Stafford, Cardamone, and
  Mazumdar]{Stafford2007}
C.~A. Stafford, D.~M. Cardamone and S.~Mazumdar, \emph{Nanotechnology}, 2007,
  \textbf{18}, 424014\relax
\mciteBstWouldAddEndPuncttrue
\mciteSetBstMidEndSepPunct{\mcitedefaultmidpunct}
{\mcitedefaultendpunct}{\mcitedefaultseppunct}\relax
\EndOfBibitem
\bibitem[P\"oltl \emph{et~al.}(2013)P\"oltl, Emary, and Brandes]{Poltl2013}
C.~P\"oltl, C.~Emary and T.~Brandes, \emph{Phys. Rev. B}, 2013, \textbf{87},
  045416\relax
\mciteBstWouldAddEndPuncttrue
\mciteSetBstMidEndSepPunct{\mcitedefaultmidpunct}
{\mcitedefaultendpunct}{\mcitedefaultseppunct}\relax
\EndOfBibitem
\bibitem[Seo \emph{et~al.}(2013)Seo, Choi, Lee, Kim, Chung, Sim, Umansky, and
  Mahalu]{Seo2013}
M.~Seo, H.~K. Choi, S.-Y. Lee, N.~Kim, Y.~Chung, H.-S. Sim, V.~Umansky and
  D.~Mahalu, \emph{Phys. Rev. Lett.}, 2013, \textbf{110}, 046803\relax
\mciteBstWouldAddEndPuncttrue
\mciteSetBstMidEndSepPunct{\mcitedefaultmidpunct}
{\mcitedefaultendpunct}{\mcitedefaultseppunct}\relax
\EndOfBibitem
\bibitem[Gu{\'e}don \emph{et~al.}(2012)Gu{\'e}don, Valkenier, Markussen,
  Thygesen, Hummelen, and van~der Molen]{Guedon2012}
C.~M. Gu{\'e}don, H.~Valkenier, T.~Markussen, K.~S. Thygesen, J.~C. Hummelen
  and S.~J. van~der Molen, \emph{Nature nanotechnology}, 2012, \textbf{7},
  304--309\relax
\mciteBstWouldAddEndPuncttrue
\mciteSetBstMidEndSepPunct{\mcitedefaultmidpunct}
{\mcitedefaultendpunct}{\mcitedefaultseppunct}\relax
\EndOfBibitem
\bibitem[Korkusinski \emph{et~al.}(2007)Korkusinski, Gimenez, Hawrylak,
  Gaudreau, Studenikin, and Sachrajda]{Korkusinski2007}
M.~Korkusinski, I.~P. Gimenez, P.~Hawrylak, L.~Gaudreau, S.~A. Studenikin and
  A.~S. Sachrajda, \emph{Phys. Rev. B}, 2007, \textbf{75}, 115301\relax
\mciteBstWouldAddEndPuncttrue
\mciteSetBstMidEndSepPunct{\mcitedefaultmidpunct}
{\mcitedefaultendpunct}{\mcitedefaultseppunct}\relax
\EndOfBibitem
\bibitem[Hsieh \emph{et~al.}(2012)Hsieh, Shim, and Hawrylak]{Hsieh2012}
C.-Y. Hsieh, Y.-P. Shim and P.~Hawrylak, \emph{Phys. Rev. B}, 2012,
  \textbf{85}, 085309\relax
\mciteBstWouldAddEndPuncttrue
\mciteSetBstMidEndSepPunct{\mcitedefaultmidpunct}
{\mcitedefaultendpunct}{\mcitedefaultseppunct}\relax
\EndOfBibitem
\bibitem[Meir and Wingreen(1992)]{Meir1992}
Y.~Meir and N.~S. Wingreen, \emph{Phys. Rev. Lett.}, 1992, \textbf{68},
  2512--2515\relax
\mciteBstWouldAddEndPuncttrue
\mciteSetBstMidEndSepPunct{\mcitedefaultmidpunct}
{\mcitedefaultendpunct}{\mcitedefaultseppunct}\relax
\EndOfBibitem
\bibitem[Kuo and Chang(2007)]{Kuo2007}
D.~M.-T. Kuo and Y.-C. Chang, \emph{Phys. Rev. Lett.}, 2007, \textbf{99},
  086803\relax
\mciteBstWouldAddEndPuncttrue
\mciteSetBstMidEndSepPunct{\mcitedefaultmidpunct}
{\mcitedefaultendpunct}{\mcitedefaultseppunct}\relax
\EndOfBibitem
\bibitem[Bu\l{}ka and Kostyrko(2004)]{Bulka2004}
B.~R. Bu\l{}ka and T.~Kostyrko, \emph{Phys. Rev. B}, 2004, \textbf{70},
  205333\relax
\mciteBstWouldAddEndPuncttrue
\mciteSetBstMidEndSepPunct{\mcitedefaultmidpunct}
{\mcitedefaultendpunct}{\mcitedefaultseppunct}\relax
\EndOfBibitem
\bibitem[Kuo \emph{et~al.}(2011)Kuo, Shiau, and Chang]{Kuo2011}
D.~M.-T. Kuo, S.-Y. Shiau and Y.-C. Chang, \emph{Phys. Rev. B}, 2011,
  \textbf{84}, 245303\relax
\mciteBstWouldAddEndPuncttrue
\mciteSetBstMidEndSepPunct{\mcitedefaultmidpunct}
{\mcitedefaultendpunct}{\mcitedefaultseppunct}\relax
\EndOfBibitem
\bibitem[Loss and DiVincenzo(1998)]{Loss1998}
D.~Loss and D.~P. DiVincenzo, \emph{Phys. Rev. A}, 1998, \textbf{57},
  120--126\relax
\mciteBstWouldAddEndPuncttrue
\mciteSetBstMidEndSepPunct{\mcitedefaultmidpunct}
{\mcitedefaultendpunct}{\mcitedefaultseppunct}\relax
\EndOfBibitem
\bibitem[Weymann \emph{et~al.}(2011)Weymann, Bu\l{}ka, and
  Barna\ifmmode~\acute{s}\else \'{s}\fi{}]{Weymann2011}
I.~Weymann, B.~Bu\l{}ka and J.~Barna\ifmmode~\acute{s}\else \'{s}\fi{},
  \emph{Phys. Rev. B}, 2011, \textbf{83}, 195302\relax
\mciteBstWouldAddEndPuncttrue
\mciteSetBstMidEndSepPunct{\mcitedefaultmidpunct}
{\mcitedefaultendpunct}{\mcitedefaultseppunct}\relax
\EndOfBibitem
\bibitem[Haug and Jauho(2007)]{Haug2007}
H.~Haug and A.-P. Jauho, \emph{Quantum kinetics in transport and optics of
  semiconductors}, Springer, 2007\relax
\mciteBstWouldAddEndPuncttrue
\mciteSetBstMidEndSepPunct{\mcitedefaultmidpunct}
{\mcitedefaultendpunct}{\mcitedefaultseppunct}\relax
\EndOfBibitem
\bibitem[Goldhaber-Gordon \emph{et~al.}(1998)Goldhaber-Gordon, Shtrikman,
  Mahalu, Abusch-Magder, Meirav, and Kastner]{Goldhaber1998}
D.~Goldhaber-Gordon, H.~Shtrikman, D.~Mahalu, D.~Abusch-Magder, U.~Meirav and
  M.~A. Kastner, \emph{Nature}, 1998, \textbf{391}, 156--159\relax
\mciteBstWouldAddEndPuncttrue
\mciteSetBstMidEndSepPunct{\mcitedefaultmidpunct}
{\mcitedefaultendpunct}{\mcitedefaultseppunct}\relax
\EndOfBibitem
\bibitem[Numata \emph{et~al.}(2009)Numata, Nisikawa, Oguri, and
  Hewson]{Numata2009}
T.~Numata, Y.~Nisikawa, A.~Oguri and A.~C. Hewson, \emph{Phys. Rev. B}, 2009,
  \textbf{80}, 155330\relax
\mciteBstWouldAddEndPuncttrue
\mciteSetBstMidEndSepPunct{\mcitedefaultmidpunct}
{\mcitedefaultendpunct}{\mcitedefaultseppunct}\relax
\EndOfBibitem
\bibitem[Michaelis \emph{et~al.}(2006)Michaelis, Emary, and
  Beenakker]{Michaelis2006}
B.~Michaelis, C.~Emary and C.~Beenakker, \emph{EPL (Europhysics Letters)},
  2006, \textbf{73}, 677\relax
\mciteBstWouldAddEndPuncttrue
\mciteSetBstMidEndSepPunct{\mcitedefaultmidpunct}
{\mcitedefaultendpunct}{\mcitedefaultseppunct}\relax
\EndOfBibitem
\bibitem[Emary(2007)]{Emary2007}
C.~Emary, \emph{Phys. Rev. B}, 2007, \textbf{76}, 245319\relax
\mciteBstWouldAddEndPuncttrue
\mciteSetBstMidEndSepPunct{\mcitedefaultmidpunct}
{\mcitedefaultendpunct}{\mcitedefaultseppunct}\relax
\EndOfBibitem
\bibitem[Kuo and Chang(2014)]{Kuo2014}
D.~M.~T. Kuo and Y.-C. Chang, \emph{Phys. Rev. B}, 2014, \textbf{89},
  115416\relax
\mciteBstWouldAddEndPuncttrue
\mciteSetBstMidEndSepPunct{\mcitedefaultmidpunct}
{\mcitedefaultendpunct}{\mcitedefaultseppunct}\relax
\EndOfBibitem
\bibitem[Bauer \emph{et~al.}(2013)Bauer, Heyder, Schubert, Borowsky, Taubert,
  Bruognolo, Schuh, Wegscheider, von Delft, and Ludwig]{Bauer2013}
F.~Bauer, J.~Heyder, E.~Schubert, D.~Borowsky, D.~Taubert, B.~Bruognolo,
  D.~Schuh, W.~Wegscheider, J.~von Delft and S.~Ludwig, \emph{Nature}, 2013,
  \textbf{501}, 73--78\relax
\mciteBstWouldAddEndPuncttrue
\mciteSetBstMidEndSepPunct{\mcitedefaultmidpunct}
{\mcitedefaultendpunct}{\mcitedefaultseppunct}\relax
\EndOfBibitem
\bibitem[Andergassen(2013)]{Andergassen2013}
S.~Andergassen, \emph{Nature}, 2013, \textbf{495}, 321--322\relax
\mciteBstWouldAddEndPuncttrue
\mciteSetBstMidEndSepPunct{\mcitedefaultmidpunct}
{\mcitedefaultendpunct}{\mcitedefaultseppunct}\relax
\EndOfBibitem
\end{mcitethebibliography}
\bibliographystyle{rsc} 
}

\end{document}